\documentclass{article}
\usepackage{graphics,color,amssymb,enumerate,epsfig}

\newcommand{\foutje}[1]{\marginpar{\tt\tiny #1}}
\renewcommand{\foutje}[1]{}

\newcommand{\fig}[1]{Fig.\ \ref{#1}}
\newcommand{\BA}{Barab\' asi-Albert}

\begin{document}
\begin{center}
{\Large\bf A network model with structured nodes}\\[2mm]
Pierluigi Frisco\\
School of Math. and Comp. Sciences, Heriot-Watt University,\\
EH14 4AS Edinburgh, UK,\\
{\tt P.Frisco@hw.ac.uk}
\end{center}

\begin{abstract}
We present a network model in which words over a specific alphabet, called {\it structures}, are associated to each node and undirected edges are added depending on some distance between different structures.

It is shown that this model can generate, without the use of preferential attachment or any other heuristic, networks with topological features similar to biological networks: power law degree distribution, clustering coefficient independent from the network size, etc.

Specific biological networks ({\it C. Elegans} neural network and {\it E. Coli} protein-protein interaction network) are replicated using this model.
\end{abstract}

\section{Introduction}
In the last years mathematical and computer science (CS) concepts and methodologies and have been successfully used in Biology. 
This fascinating and fruitful combination of these disciplines has clear advantages for both of them. 
When biological phenomena are regarded as information processes, then they can be studied using mathematical and CS tools and concepts. This gives to Biology new ways to approach problems, solutions to them and this deepens the understanding of biological processes. At the same time CS enriches itself with new ways to define and study information process while Mathematics enriches itself with new concepts and theories.

In the last decade several studies (\cite{alonBook,motifs,JTeo00,superfamilies}) showed the importance of the topology of biological networks. 
These results proved that biological networks are composed of motifs, that biological networks with specific functions have an abundance of certain motifs instead of others, that the number of edges for the node in the network follows specific laws, etc.

More than studying the features of empirical networks, it is also important to have algorithms able to generate networks with the same features of empirical ones.
This kind of algorithms, called {\it models}, are an invaluable help in the generation of artificial networks and they provide insights on how certain features of complex empirical networks arise from the construction rules present in the model.

Examples of such procedures are: the {\it Erd\"os-R\'enyi} model \cite{ER59}, the {\it Watts-Strogatz} model \cite{WS98} and the {\it Bar\' abasi-Albert} model \cite{BA99} and its variants \cite{AB02,RSM02}.
The E-R model allows to generate random networks able to reproduce the small-world property (short path from any node to any other node in the network) but they fail to account for the local clustering characterising many empirical networks. Both these properties are captures by the W-S model, but unfortunately it does not capture the inhomogeneous degree distribution found in many empirical networks.
The B-A model can overcome these limitations and gives rise to the degree distribution. 
This degree distribution is obtained using preferential attachment: the probability for a node to receive an edge depends on the number of edges the node already has.
The original B-A model does not capture the independence of the clustering coefficient from the size (number of nodes) of a network.
This feature is captured by a variant \cite{RSM02} of this model in which heuristics (replication of networks) are used.\\

The present study originates from the wish to create a network model able to reproduce biological networks without the use of heuristics.
Despite the very many successful applications of the B-A model, it was not clear to us how preferential attachment could have been present in the evolution of, say, gene networks.
Why a gene with many interactions is more likely to get even more interactions than a gene with few interactions?
How can a new added gene ``know'' what are the genes with more interactions?
In this respect, we believe that preferential attachment capture the overall effect of something more basic present in the evolution of biological networks.

The network model introduced and studied in the present paper tries to capture some basic features present in the evolution of biological networks: network growth, node structure and distance between node structures.

The node structure represents, for instance, the DNA sequence in genes, proteins' secondary structure, the personality features in humans, etc.
The distance between nodes represents, for instance, the fact that proteins will interact if their tertiary structure (which depends on their secondary structure) allows it, or that two humans will be friends if the treats of their personality are somehow close.

In the following we present the model with structured nodes (Section \ref{sec:descr}), we analyse it (Section \ref{sec:compBA}) and we use it to generate specific biological networks (Section \ref{sec:reprNet}).
The paper ends with a discussion section (Section \ref{sec:fr}).
Supplementary material (further technical details, generated networks, program implementing the proposed network model, etc.) is present at \cite{myDownload}.

\section{Description of the model}
\label{sec:descr}
The {\it network model with structured node} (SN model) is such that each node in the network has a {\it structure}: a word over a specified alphabet.
Given initial nodes have different structure.
Nodes are added to the network one by one.
Each new node has a structure given by the modification of a randomly chosen structure already present in the network.
If the structure of the new node is already present in the network, then the new node is not added (that is, in the network all nodes have different structure).
If the structure of the new node is not present in the network and the new node has no edge with the existing nodes, then the new node is not added (that is, isolated nodes are not allowed).
Undirected edges are added to the network depending on a given distance between node structures.
This process is repeated until the network reaches a given number of nodes.
A simple example follows.

Let us assume that the {\it alphabet} is {\tt \{A, B, C\}} and that the network contains only one {\it initial node} with structure {\tt ABCABC}.
Edges between nodes are added only if the {\it Hamming distance} \cite{wikip09} between the structures of the nodes is at most 1.

A node can be added to the network by {\it mutating} one symbol in the structure of an existing node.
For instance, the node {\tt ABBABC} can be obtained mutating the third symbol of {\tt ABCABC}. 
An edge is added between the two nodes (they only differ in one symbol).

A third node can be added to the network by {\it adding} one symbol to the structure of a randomly selected existing node.
For instance, the node {\tt ABBABBC} can be obtained adding a {\tt B} between {\tt B} and {\tt C} in node {\tt ABBABC}.
An edge is added between the new node and {\tt ABBABC} (when computing the distance between two structures exceeding symbols in the longer structure are disregarded).
No edge is added between the new node and {\tt ABCABC} because there are 2 differences in their first 6 characters.

The structure {\tt ABCBC} can be obtained from {\tt ABCABC} {\it deleting} the second {\tt B}.
The node with this new structure does not become part of the network as no edge has been added (the distance between {\tt ABCBC} and the other structures present in the network is bigger than 1).

The structure {\tt ABBBBABBC} can be obtained from {\tt ABBABBC} {\it duplicating} the second and third {\tt B}.
The node with this new structure does not become part of the network as no edge has been added.

Input parameters define the probabilities to mutate, add, delete and duplicate node structures and their values has to sum up to 1.

We also used a Hamming distance in which the comparison between symbols considers groups of consecutive symbols. 
The order of the symbols present in each such group is irrelevant to the distance.
For instance, let us consider the two structures {\tt ABBABC} and {\tt BABCAB}.
If the {\it unit distance} is 1 (i.e., symbols are compared one by one), then the distance between the two structures is 5 as the only matching symbol is the {\tt B} in the third position.
If the unit distance is 2 (i.e., pairs of symbols are compared), then the distance between the two structures is 2.
This is because the first two pairs are considered equal ({\tt AB} and {\tt BA} differ only in the order of the symbols), and the other two pairs are different in the symbols they contain.
If the unit distance is 3 (i.e., triplets of symbols are compared), then the distance between the two structures is 0.
This is because the first triples are considered equal ({\tt ABB} and {\tt BAB} differ only in the order of symbols) as well as the second triple ({\tt ABC} and {\tt CAB} differ only in the order of symbols).

An edge between two nodes is present only if their distance is smaller/equal than the value of the input parameter {\it maximum distance}.

When unit distance is bigger than 1, then it is possible to have a {\it file matches} indicating how the different groups of symbols can be matched to eachother.
In other words, a file matches behaves as the genetic code: it denotes which tuples of symbols have to be regarded as equal (in the same way different codons translate in the same amino acid).
For instance, let unit distance be 2, the alphabet be {\tt \{A, B\}}, and the file matches be:\\
{\tt 
AB =\\
BA =\\
AA = BB\\
BB = AA\\
}
With this file matches, the strings {\tt ABBB} and {\tt ABAA} have distance 0.
This is because the first pair ({\tt AB}) is the same in both strings, while the second pair ({\tt BB} and {\tt AA}) is defined by the file matches to be equal.
Without the file matches, the two string have distance 1 (due to the second pair).

We call {\it instance} a set of input parameters.
The complete list of input parameters together with their description can be found in the user manual of the program implementing the SN model \cite{myDownload}.

\section{Analysis of the model}
\label{sec:compBA}
We assessed our network model over the following network topological features \cite{ABN08}.
Given an undirected network $G$ with $N$ nodes and $k$ edges we denote by 
$\langle k\rangle$ the {\it average degree}, 
by $L$ the {\it average path length},
by $C$ the {\it average clustering coefficient},
by $P(k)$ the {\it degree distribution} and
by $C(k)$ the {\it clustering coefficient distribution}.

We also considered the:
\begin{description}
\item 3-node {\it motifs distribution}, that is the number (normalised to 1) of triples of nodes having no edge, only 1 edge, only 2 edges and 3 edges between themselves;
\item {\it path length distribution}, denoted by $PL(\ell)$, relating the number (normalised to one) of paths having a certain length $\ell$;
\item {\it heterogeneity index}, denoted by $\rho(G)$ (where $G$ is the network), a new formulation of Randi\' c index introduced in \cite{Es10-1,Es10-2}. In \cite{Es10-1,Es10-2} it is also shown that the \BA$\;$model is not able to generate network with a heterogeneity index as high as the one found in biological networks.
\end{description}

We compared the network generated by an instance our the SN model with the network generated by the \BA$\;$model (our implementation of this model is based on the Fortran implementation present at \cite{BAfortran}).
For this purpose we run the \BA$\;$model starting with a clique of 6 nodes and adding 6 edges for each new added node.
We also run the following instance of our network model: initial node {\tt ABCDEFGHILMN}, alphabet {\tt {A, ..., T}}, probability to mutate 1 (which implies that the length of the node structures is equal to the one of the initial node), Hamming distance having unit distance 2 and maximum distance 2.
We run these simulation for 3000 iterations storing the resulting intermediate networks every 500 iterations.
These tests run 100 times for different random seeds.

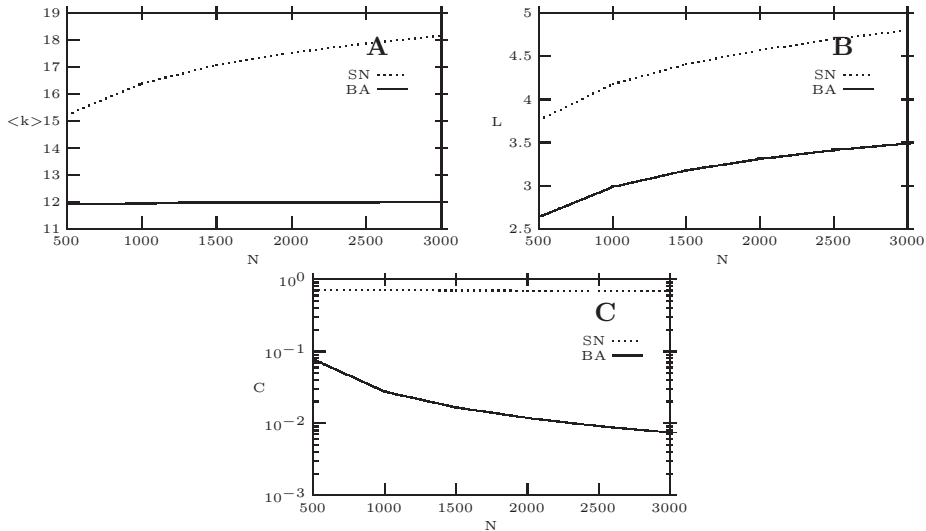
\begin{figure}[h]
\begin{center}
\setlength{\unitlength}{0.110900pt}
\ifx\plotpoint\undefined\newsavebox{\plotpoint}\fi
\sbox{\plotpoint}{\rule[-0.200pt]{0.400pt}{0.400pt}}%
\begin{picture}(1500,900)(0,0)
\sbox{\plotpoint}{\rule[-0.200pt]{0.400pt}{0.400pt}}%
\put(161.0,123.0){\rule[-0.200pt]{4.818pt}{0.400pt}}
\put(141,123){\makebox(0,0)[r]{\tiny 11}}
\put(1419.0,123.0){\rule[-0.200pt]{4.818pt}{0.400pt}}
\put(161.0,215.0){\rule[-0.200pt]{4.818pt}{0.400pt}}
\put(141,215){\makebox(0,0)[r]{\tiny 12}}
\put(1419.0,215.0){\rule[-0.200pt]{4.818pt}{0.400pt}}
\put(161.0,307.0){\rule[-0.200pt]{4.818pt}{0.400pt}}
\put(141,307){\makebox(0,0)[r]{\tiny 13}}
\put(1419.0,307.0){\rule[-0.200pt]{4.818pt}{0.400pt}}
\put(161.0,399.0){\rule[-0.200pt]{4.818pt}{0.400pt}}
\put(141,399){\makebox(0,0)[r]{\tiny 14}}
\put(1419.0,399.0){\rule[-0.200pt]{4.818pt}{0.400pt}}
\put(161.0,492.0){\rule[-0.200pt]{4.818pt}{0.400pt}}
\put(141,492){\makebox(0,0)[r]{\tiny 15}}
\put(1419.0,492.0){\rule[-0.200pt]{4.818pt}{0.400pt}}
\put(161.0,584.0){\rule[-0.200pt]{4.818pt}{0.400pt}}
\put(141,584){\makebox(0,0)[r]{\tiny 16}}
\put(1419.0,584.0){\rule[-0.200pt]{4.818pt}{0.400pt}}
\put(161.0,676.0){\rule[-0.200pt]{4.818pt}{0.400pt}}
\put(141,676){\makebox(0,0)[r]{\tiny 17}}
\put(1419.0,676.0){\rule[-0.200pt]{4.818pt}{0.400pt}}
\put(161.0,768.0){\rule[-0.200pt]{4.818pt}{0.400pt}}
\put(141,768){\makebox(0,0)[r]{\tiny 18}}
\put(1419.0,768.0){\rule[-0.200pt]{4.818pt}{0.400pt}}
\put(161.0,860.0){\rule[-0.200pt]{4.818pt}{0.400pt}}
\put(141,860){\makebox(0,0)[r]{\tiny 19}}
\put(1419.0,860.0){\rule[-0.200pt]{4.818pt}{0.400pt}}
\put(161.0,123.0){\rule[-0.200pt]{0.400pt}{4.818pt}}
\put(161,82){\makebox(0,0){\tiny 500}}
\put(161.0,840.0){\rule[-0.200pt]{0.400pt}{4.818pt}}
\put(417.0,123.0){\rule[-0.200pt]{0.400pt}{4.818pt}}
\put(417,82){\makebox(0,0){\tiny 1000}}
\put(417.0,840.0){\rule[-0.200pt]{0.400pt}{4.818pt}}
\put(672.0,123.0){\rule[-0.200pt]{0.400pt}{4.818pt}}
\put(672,82){\makebox(0,0){\tiny 1500}}
\put(672.0,840.0){\rule[-0.200pt]{0.400pt}{4.818pt}}
\put(928.0,123.0){\rule[-0.200pt]{0.400pt}{4.818pt}}
\put(928,82){\makebox(0,0){\tiny 2000}}
\put(928.0,840.0){\rule[-0.200pt]{0.400pt}{4.818pt}}
\put(1183.0,123.0){\rule[-0.200pt]{0.400pt}{4.818pt}}
\put(1183,82){\makebox(0,0){\tiny 2500}}
\put(1183.0,840.0){\rule[-0.200pt]{0.400pt}{4.818pt}}
\put(1439.0,123.0){\rule[-0.200pt]{0.400pt}{4.818pt}}
\put(1439,82){\makebox(0,0){\tiny 3000}}
\put(1439.0,840.0){\rule[-0.200pt]{0.400pt}{4.818pt}}
\put(20,491){\makebox(0,0){\tiny <k>}}
\put(800,21){\makebox(0,0){\tiny N}}
\put(1183,749){\makebox(0,0)[l]{\bf A}}
\put(1200,600){\makebox(0,0)[r]{\tiny BA}}
\put(1224.0,600.0){\rule[-0.200pt]{11.090pt}{0.400pt}}
\put(161,207){\usebox{\plotpoint}}
\multiput(161.00,207.60)(37.329,0.468){5}{\rule{25.700pt}{0.113pt}}
\multiput(161.00,206.17)(202.658,4.000){2}{\rule{12.850pt}{0.400pt}}
\put(417,210.67){\rule{61.430pt}{0.400pt}}
\multiput(417.00,210.17)(127.500,1.000){2}{\rule{30.715pt}{0.400pt}}
\put(672,211.67){\rule{61.670pt}{0.400pt}}
\multiput(672.00,211.17)(128.000,1.000){2}{\rule{30.835pt}{0.400pt}}
\put(928,212.67){\rule{55.430pt}{0.400pt}}
\put(1200,650){\makebox(0,0)[r]{\tiny SN}}
\multiput(1224,650)(20.756,0.000){5}{\usebox{\plotpoint}}
\put(1324,4780){\usebox{\plotpoint}}
\put(161,512){\usebox{\plotpoint}}
\multiput(161,512)(19.123,8.068){14}{\usebox{\plotpoint}}
\multiput(417,620)(20.150,4.978){13}{\usebox{\plotpoint}}
\multiput(672,683)(20.482,3.360){12}{\usebox{\plotpoint}}
\multiput(928,725)(20.594,2.584){12}{\usebox{\plotpoint}}
\multiput(1183,757)(20.649,2.097){13}{\usebox{\plotpoint}}
\put(1439,783){\usebox{\plotpoint}}
\put(161.0,123.0){\rule[-0.200pt]{140.870pt}{0.400pt}}
\put(1439.0,123.0){\rule[-0.200pt]{0.400pt}{82.543pt}}
\put(161.0,860.0){\rule[-0.200pt]{140.870pt}{0.400pt}}
\put(161.0,123.0){\rule[-0.200pt]{0.400pt}{82.543pt}}
\end{picture}
\setlength{\unitlength}{0.110900pt}
\ifx\plotpoint\undefined\newsavebox{\plotpoint}\fi
\sbox{\plotpoint}{\rule[-0.200pt]{0.400pt}{0.400pt}}%
\begin{picture}(1500,900)(0,0)
\sbox{\plotpoint}{\rule[-0.200pt]{0.400pt}{0.400pt}}%
\put(181.0,123.0){\rule[-0.200pt]{4.818pt}{0.400pt}}
\put(161,123){\makebox(0,0)[r]{\tiny 2.5}}
\put(1419.0,123.0){\rule[-0.200pt]{4.818pt}{0.400pt}}
\put(181.0,270.0){\rule[-0.200pt]{4.818pt}{0.400pt}}
\put(161,270){\makebox(0,0)[r]{\tiny 3}}
\put(1419.0,270.0){\rule[-0.200pt]{4.818pt}{0.400pt}}
\put(181.0,418.0){\rule[-0.200pt]{4.818pt}{0.400pt}}
\put(161,418){\makebox(0,0)[r]{\tiny 3.5}}
\put(1419.0,418.0){\rule[-0.200pt]{4.818pt}{0.400pt}}
\put(181.0,565.0){\rule[-0.200pt]{4.818pt}{0.400pt}}
\put(161,565){\makebox(0,0)[r]{\tiny 4}}
\put(1419.0,565.0){\rule[-0.200pt]{4.818pt}{0.400pt}}
\put(181.0,713.0){\rule[-0.200pt]{4.818pt}{0.400pt}}
\put(161,713){\makebox(0,0)[r]{\tiny 4.5}}
\put(1419.0,713.0){\rule[-0.200pt]{4.818pt}{0.400pt}}
\put(181.0,860.0){\rule[-0.200pt]{4.818pt}{0.400pt}}
\put(161,860){\makebox(0,0)[r]{\tiny 5}}
\put(1419.0,860.0){\rule[-0.200pt]{4.818pt}{0.400pt}}
\put(181.0,123.0){\rule[-0.200pt]{0.400pt}{4.818pt}}
\put(181,82){\makebox(0,0){\tiny 500}}
\put(181.0,840.0){\rule[-0.200pt]{0.400pt}{4.818pt}}
\put(433.0,123.0){\rule[-0.200pt]{0.400pt}{4.818pt}}
\put(433,82){\makebox(0,0){\tiny 1000}}
\put(433.0,840.0){\rule[-0.200pt]{0.400pt}{4.818pt}}
\put(684.0,123.0){\rule[-0.200pt]{0.400pt}{4.818pt}}
\put(684,82){\makebox(0,0){\tiny 1500}}
\put(684.0,840.0){\rule[-0.200pt]{0.400pt}{4.818pt}}
\put(936.0,123.0){\rule[-0.200pt]{0.400pt}{4.818pt}}
\put(936,82){\makebox(0,0){\tiny 2000}}
\put(936.0,840.0){\rule[-0.200pt]{0.400pt}{4.818pt}}
\put(1187.0,123.0){\rule[-0.200pt]{0.400pt}{4.818pt}}
\put(1187,82){\makebox(0,0){\tiny 2500}}
\put(1187.0,840.0){\rule[-0.200pt]{0.400pt}{4.818pt}}
\put(1439.0,123.0){\rule[-0.200pt]{0.400pt}{4.818pt}}
\put(1439,82){\makebox(0,0){\tiny 3000}}
\put(1439.0,840.0){\rule[-0.200pt]{0.400pt}{4.818pt}}
\put(40,491){\makebox(0,0){\tiny L}}
\put(810,21){\makebox(0,0){\tiny N}}
\put(1183,749){\makebox(0,0)[l]{\bf B}}
\put(1200,600){\makebox(0,0)[r]{\tiny BA}}
\put(1224.0,600.0){\rule[-0.200pt]{11.090pt}{0.400pt}}
\put(181,164){\usebox{\plotpoint}}
\multiput(181.00,164.58)(1.225,0.499){203}{\rule{1.079pt}{0.120pt}}
\multiput(181.00,163.17)(249.761,103.000){2}{\rule{0.539pt}{0.400pt}}
\multiput(433.00,267.58)(2.212,0.499){111}{\rule{1.861pt}{0.120pt}}
\multiput(433.00,266.17)(247.137,57.000){2}{\rule{0.931pt}{0.400pt}}
\multiput(684.00,324.58)(3.255,0.498){75}{\rule{2.685pt}{0.120pt}}
\multiput(684.00,323.17)(246.428,39.000){2}{\rule{1.342pt}{0.400pt}}
\multiput(936.00,363.58)(4.227,0.497){57}{\rule{3.447pt}{0.120pt}}
\multiput(936.00,362.17)(243.846,30.000){2}{\rule{1.723pt}{0.400pt}}
\multiput(1187.00,393.58)(5.557,0.496){43}{\rule{4.483pt}{0.120pt}}
\multiput(1187.00,392.17)(242.696,23.000){2}{\rule{2.241pt}{0.400pt}}
\put(1200,650){\makebox(0,0)[r]{\tiny SN}}
\multiput(1224,650)(20.756,0.000){5}{\usebox{\plotpoint}}
\put(1328,1202){\usebox{\plotpoint}}
\put(181,494){\usebox{\plotpoint}}
\multiput(181,494)(18.623,9.164){14}{\usebox{\plotpoint}}
\multiput(433,618)(20.033,5.427){13}{\usebox{\plotpoint}}
\multiput(684,686)(20.389,3.884){12}{\usebox{\plotpoint}}
\multiput(936,734)(20.534,3.027){12}{\usebox{\plotpoint}}
\multiput(1187,771)(20.600,2.534){12}{\usebox{\plotpoint}}
\put(1439,802){\usebox{\plotpoint}}
\put(181.0,123.0){\rule[-0.200pt]{140.052pt}{0.400pt}}
\put(1439.0,123.0){\rule[-0.200pt]{0.400pt}{82.543pt}}
\put(181.0,860.0){\rule[-0.200pt]{140.052pt}{0.400pt}}
\put(181.0,123.0){\rule[-0.200pt]{0.400pt}{82.543pt}}
\end{picture}
\setlength{\unitlength}{0.110900pt}
\ifx\plotpoint\undefined\newsavebox{\plotpoint}\fi
\sbox{\plotpoint}{\rule[-0.200pt]{0.400pt}{0.400pt}}%
\begin{picture}(1500,900)(0,0)
\sbox{\plotpoint}{\rule[-0.200pt]{0.400pt}{0.400pt}}%
\put(221.0,123.0){\rule[-0.200pt]{4.818pt}{0.400pt}}
\put(201,123){\makebox(0,0)[r]{\tiny $10^{-3}$}}
\put(1419.0,123.0){\rule[-0.200pt]{4.818pt}{0.400pt}}
\put(221.0,197.0){\rule[-0.200pt]{2.409pt}{0.400pt}}
\put(1429.0,197.0){\rule[-0.200pt]{2.409pt}{0.400pt}}
\put(221.0,240.0){\rule[-0.200pt]{2.409pt}{0.400pt}}
\put(1429.0,240.0){\rule[-0.200pt]{2.409pt}{0.400pt}}
\put(221.0,271.0){\rule[-0.200pt]{2.409pt}{0.400pt}}
\put(1429.0,271.0){\rule[-0.200pt]{2.409pt}{0.400pt}}
\put(221.0,295.0){\rule[-0.200pt]{2.409pt}{0.400pt}}
\put(1429.0,295.0){\rule[-0.200pt]{2.409pt}{0.400pt}}
\put(221.0,314.0){\rule[-0.200pt]{2.409pt}{0.400pt}}
\put(1429.0,314.0){\rule[-0.200pt]{2.409pt}{0.400pt}}
\put(221.0,331.0){\rule[-0.200pt]{2.409pt}{0.400pt}}
\put(1429.0,331.0){\rule[-0.200pt]{2.409pt}{0.400pt}}
\put(221.0,345.0){\rule[-0.200pt]{2.409pt}{0.400pt}}
\put(1429.0,345.0){\rule[-0.200pt]{2.409pt}{0.400pt}}
\put(221.0,357.0){\rule[-0.200pt]{2.409pt}{0.400pt}}
\put(1429.0,357.0){\rule[-0.200pt]{2.409pt}{0.400pt}}
\put(221.0,369.0){\rule[-0.200pt]{4.818pt}{0.400pt}}
\put(201,369){\makebox(0,0)[r]{\tiny $10^{-2}$}}
\put(1419.0,369.0){\rule[-0.200pt]{4.818pt}{0.400pt}}
\put(221.0,443.0){\rule[-0.200pt]{2.409pt}{0.400pt}}
\put(1429.0,443.0){\rule[-0.200pt]{2.409pt}{0.400pt}}
\put(221.0,486.0){\rule[-0.200pt]{2.409pt}{0.400pt}}
\put(1429.0,486.0){\rule[-0.200pt]{2.409pt}{0.400pt}}
\put(221.0,517.0){\rule[-0.200pt]{2.409pt}{0.400pt}}
\put(1429.0,517.0){\rule[-0.200pt]{2.409pt}{0.400pt}}
\put(221.0,540.0){\rule[-0.200pt]{2.409pt}{0.400pt}}
\put(1429.0,540.0){\rule[-0.200pt]{2.409pt}{0.400pt}}
\put(221.0,560.0){\rule[-0.200pt]{2.409pt}{0.400pt}}
\put(1429.0,560.0){\rule[-0.200pt]{2.409pt}{0.400pt}}
\put(221.0,576.0){\rule[-0.200pt]{2.409pt}{0.400pt}}
\put(1429.0,576.0){\rule[-0.200pt]{2.409pt}{0.400pt}}
\put(221.0,591.0){\rule[-0.200pt]{2.409pt}{0.400pt}}
\put(1429.0,591.0){\rule[-0.200pt]{2.409pt}{0.400pt}}
\put(221.0,603.0){\rule[-0.200pt]{2.409pt}{0.400pt}}
\put(1429.0,603.0){\rule[-0.200pt]{2.409pt}{0.400pt}}
\put(221.0,614.0){\rule[-0.200pt]{4.818pt}{0.400pt}}
\put(201,614){\makebox(0,0)[r]{\tiny $10^{-1}$}}
\put(1419.0,614.0){\rule[-0.200pt]{4.818pt}{0.400pt}}
\put(221.0,688.0){\rule[-0.200pt]{2.409pt}{0.400pt}}
\put(1429.0,688.0){\rule[-0.200pt]{2.409pt}{0.400pt}}
\put(221.0,732.0){\rule[-0.200pt]{2.409pt}{0.400pt}}
\put(1429.0,732.0){\rule[-0.200pt]{2.409pt}{0.400pt}}
\put(221.0,762.0){\rule[-0.200pt]{2.409pt}{0.400pt}}
\put(1429.0,762.0){\rule[-0.200pt]{2.409pt}{0.400pt}}
\put(221.0,786.0){\rule[-0.200pt]{2.409pt}{0.400pt}}
\put(1429.0,786.0){\rule[-0.200pt]{2.409pt}{0.400pt}}
\put(221.0,805.0){\rule[-0.200pt]{2.409pt}{0.400pt}}
\put(1429.0,805.0){\rule[-0.200pt]{2.409pt}{0.400pt}}
\put(221.0,822.0){\rule[-0.200pt]{2.409pt}{0.400pt}}
\put(1429.0,822.0){\rule[-0.200pt]{2.409pt}{0.400pt}}
\put(221.0,836.0){\rule[-0.200pt]{2.409pt}{0.400pt}}
\put(1429.0,836.0){\rule[-0.200pt]{2.409pt}{0.400pt}}
\put(221.0,849.0){\rule[-0.200pt]{2.409pt}{0.400pt}}
\put(1429.0,849.0){\rule[-0.200pt]{2.409pt}{0.400pt}}
\put(221.0,860.0){\rule[-0.200pt]{4.818pt}{0.400pt}}
\put(201,860){\makebox(0,0)[r]{\tiny $10^0$}}
\put(1419.0,860.0){\rule[-0.200pt]{4.818pt}{0.400pt}}
\put(221.0,123.0){\rule[-0.200pt]{0.400pt}{4.818pt}}
\put(221,82){\makebox(0,0){\tiny 500}}
\put(221.0,840.0){\rule[-0.200pt]{0.400pt}{4.818pt}}
\put(465.0,123.0){\rule[-0.200pt]{0.400pt}{4.818pt}}
\put(465,82){\makebox(0,0){\tiny 1000}}
\put(465.0,840.0){\rule[-0.200pt]{0.400pt}{4.818pt}}
\put(708.0,123.0){\rule[-0.200pt]{0.400pt}{4.818pt}}
\put(708,82){\makebox(0,0){\tiny 1500}}
\put(708.0,840.0){\rule[-0.200pt]{0.400pt}{4.818pt}}
\put(952.0,123.0){\rule[-0.200pt]{0.400pt}{4.818pt}}
\put(952,82){\makebox(0,0){\tiny 2000}}
\put(952.0,840.0){\rule[-0.200pt]{0.400pt}{4.818pt}}
\put(1195.0,123.0){\rule[-0.200pt]{0.400pt}{4.818pt}}
\put(1195,82){\makebox(0,0){\tiny 2500}}
\put(1195.0,840.0){\rule[-0.200pt]{0.400pt}{4.818pt}}
\put(1439.0,123.0){\rule[-0.200pt]{0.400pt}{4.818pt}}
\put(1439,82){\makebox(0,0){\tiny 3000}}
\put(1439.0,840.0){\rule[-0.200pt]{0.400pt}{4.818pt}}
\put(40,491){\makebox(0,0){\tiny C}}
\put(830,21){\makebox(0,0){\tiny N}}
\put(1183,749){\makebox(0,0)[l]{\bf C}}
\put(1224,600){\makebox(0,0)[r]{\tiny BA}}
\put(1244.0,600.0){\rule[-0.200pt]{11.090pt}{0.400pt}}
\put(221,587){\usebox{\plotpoint}}
\multiput(221.00,585.92)(1.101,-0.499){219}{\rule{0.979pt}{0.120pt}}
\multiput(221.00,586.17)(241.967,-111.000){2}{\rule{0.490pt}{0.400pt}}
\multiput(465.00,474.92)(2.261,-0.498){105}{\rule{1.900pt}{0.120pt}}
\multiput(465.00,475.17)(239.056,-54.000){2}{\rule{0.950pt}{0.400pt}}
\multiput(708.00,420.92)(3.417,-0.498){69}{\rule{2.811pt}{0.120pt}}
\multiput(708.00,421.17)(238.165,-36.000){2}{\rule{1.406pt}{0.400pt}}
\multiput(952.00,384.92)(4.388,-0.497){53}{\rule{3.571pt}{0.120pt}}
\multiput(952.00,385.17)(235.587,-28.000){2}{\rule{1.786pt}{0.400pt}}
\multiput(1195.00,356.92)(5.629,-0.496){41}{\rule{4.536pt}{0.120pt}}
\multiput(1195.00,357.17)(234.585,-22.000){2}{\rule{2.268pt}{0.400pt}}
\put(1220,650){\makebox(0,0)[r]{\tiny SN}}
\multiput(1244,650)(20.756,0.000){5}{\usebox{\plotpoint}}
\put(221,825){\usebox{\plotpoint}}
\multiput(221,825)(20.755,-0.085){12}{\usebox{\plotpoint}}
\multiput(465,824)(20.755,-0.085){12}{\usebox{\plotpoint}}
\multiput(708,823)(20.755,-0.085){12}{\usebox{\plotpoint}}
\multiput(952,822)(20.756,0.000){11}{\usebox{\plotpoint}}
\multiput(1195,822)(20.756,0.000){12}{\usebox{\plotpoint}}
\put(1439,822){\usebox{\plotpoint}}
\put(221.0,123.0){\rule[-0.200pt]{135.416pt}{0.400pt}}
\put(1439.0,123.0){\rule[-0.200pt]{0.400pt}{82.543pt}}
\put(221.0,860.0){\rule[-0.200pt]{135.416pt}{0.400pt}}
\put(221.0,123.0){\rule[-0.200pt]{0.400pt}{82.543pt}}
\end{picture}
\caption{{\bf (A)} average degree, {\bf (B)} average path length and {\bf (C)} average clustering coefficient of a growing \BA$\;$model network (BA) and a growing SN model network.}
\label{fig:BAres}
\end{center}
\end{figure}

Figure \ref{fig:BAres} shows how the average degree, average path length and average clustering coefficient change in the \BA$\;$model and in the SN model.

The average path length follows the same curve in both models and the average degree slowly grows in the SN model while it remain constant in the \BA$\;$model.
The major difference is present in the clustering coefficient: in remains constant in the SN model while it decreases fast in the \BA$\;$model.
It is known that empirical networks have a clustering coefficient independent from their size and in \cite{RSMeo02} a variant of the \BA$\;$model generating networks with a power law degree distribution and a clustering coefficient independent from the size of the network was presented.
The motif distribution was similar in both model (data not shown).

It is well known that the \BA$\;$model generates networks with a degree distribution following a power law $P(k) \sim k^{-\gamma}$.
The same holds true for the considered instance of the SN model (this is not true for all instances of the SN  model).

\begin{figure}[h]
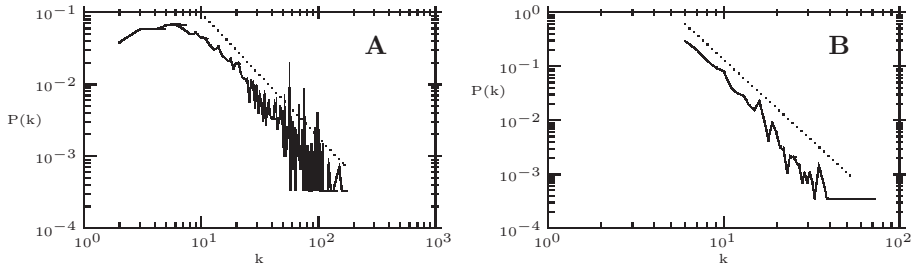

\begin{center}
\input{./SNedgesDistr_modified.tex}
\input{./SNedgesDistr_55000_modified.tex}
\caption{Degree distribution of two instances of the SN model: {\bf (A)}: keeping all nodes (final trend having slope $\gamma = -1.72$), {\bf (B)} removing nodes with less than 5 edges together with their edges (trend having slope $\gamma = -2.98$).}
\label{fig:SNedges}
\end{center}
\end{figure}

In both models the exponent of the power law does not change during growth.
Anyhow, in the considered instance, the degree distribution of the networks generated by the SN model is not following a power law in its initial phases. 
This is shown in \fig{fig:SNedges}A where it can be seen that only after $k=5$ the degree distribution follows a power law.
This difference with the \BA$\;$model is mainly due to the fact that in the \BA$\;$model each new added node has a fixed number of edges (6 in the case considered by us), while this request for a minimum number of edges is not present in the SN model.

We run another instance of the SN model for 55000 iterations and then we let all nodes having less than 5 edges to be removed from the generated networks together with their edges.
The resulting networks, having around 3000 nodes, have a power law degree distribution \fig{fig:SNedges}B.

As already said, in the SN model new nodes are added to the network first selecting the structure of a node already in the network, then changing it and finally adding it to the network.
In order to study if the selection criteria had an influence on the generated network we run a variant of the SN model.
In this variant new nodes are added all at once and their structure is a modification of the structure of the nodes initially present.
So, the difference is that only the initial structures are used as template for the new added structures.

Network generated in this way showed a very low average clustering coefficient (around 0.002) and their degree distribution did not follow a power low (see supplementary material \cite{myDownload}).

From this we conclude that, in the SN model, the incremental addition of new nodes to the network (as opposed to the addition of the nodes all at once) is a necessary element in order to have a power low degree distribution and a high average clustering coefficient.

\section{Reproduced networks}
\label{sec:reprNet}
Different instances allow the SN model to generate networks with different topological features.
In this section we describe how this model can generate networks having topological features similar to the ones of some empirical networks.
We followed a trial-and-error approach in order to find the input parameters to `fit the networks': manually testing different instances until a `sufficiently good' result was found.
We are confident that a different approach, either analytical or based on heuristics (i.e., evolutionary algorithms \cite{debBook}), could lead to better results.

All the (input and output) files related to the reproduced network described in the following are available from \cite{myDownload}.
We successfully generated networks having strong similarities with the MRSA gene network (data not shown).\\

\noindent
{\bf C. Elegans neural network}.
In \cite{WS98} it is reported that the neural network of {\it C. Elegans} has 282 nodes (neurons), an average degree of 14, an average path length of 2.65 and an average clustering coefficient of 0.28.
We were not able to retrieve a description of this network, so we only tried to match the just given network topological features.

Using the described network model we run tests having: {\tt \{A, T\}} as alphabet, {\tt ATATATATATAT} as structure of the only initial node, probability to mutate equal to 1, unit distance equal to 2 and nodes have a common edge only if their Hamming distance is smaller than 1.
The generated networks with 282 nodes have (average on 100 tests): 13.94 as average degree, 3.61 as average path length and 0.3 as average clustering coefficient.

The 100 generated networks have a low variance on these values: 61$\%$ have at most 10$\%$ discrepancy from the average results (see supplementary material \cite{myDownload}).\\

\noindent
{\bf {\it E. Coli} protein-protein interaction network}.
In \cite{Betal05} the protein-protein interaction (PPI) network of {\it E. Coli} was published.
The biggest connected component of this network consist of 230 nodes, it has an average degree of 6.04, an average path length of 3.78, an average clustering coefficient of 0.22 and a heterogeneity index of 0.24.

Using the SN model we run tests having:
{\tt \{A, T, C\}} as alphabet,
{\tt ATCATCTCATCACT} as structure of the only initial node,
probability to mutate equal to 0.4,
probability to duplicate equal to 0.6,
unit distance equal to 2,
using the file matches considered to reproduce the MRSA gene network
and nodes have a common edge only if their Hamming distance is smaller/equal than 1.

The generated networks with 230 nodes have (average over 100 tests):
6.03 as average degree,
3.85 as average path length,
0.47 as average clustering coefficient and 
0.26 as heterogeneity index.
The variance of these networks is rather big: only 3$\%$ have at most 10$\%$ dicsrepancy from the average results while only 24$\%$ have at most 20$\%$ discrepancy from the average results (see supplementary material \cite{myDownload}).
In general, we noticed that the probability to duplicate increases the heterogeneity index of a network but also increases the variance of the generated networks.

The degree distribution of the given network follows a power low with trend $\gamma = -1.06$.
The degree distribution of the networks generated by us also follow a power law but (average over 100 tests) with trend $\gamma = 0.72$.
It is remarkable that the SN model is able to generate networks with a small number of nodes having a power law as degree distribution. 

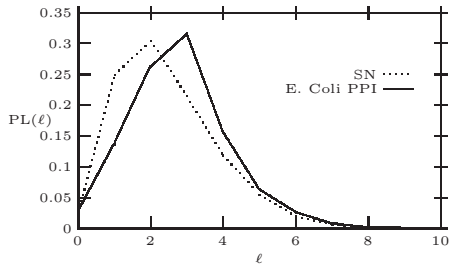
\begin{figure}[h]
\begin{center}
\setlength{\unitlength}{0.110900pt}
\ifx\plotpoint\undefined\newsavebox{\plotpoint}\fi
\sbox{\plotpoint}{\rule[-0.200pt]{0.400pt}{0.400pt}}%
\begin{picture}(1500,900)(0,0)
\sbox{\plotpoint}{\rule[-0.200pt]{0.400pt}{0.400pt}}%
\put(201.0,123.0){\rule[-0.200pt]{4.818pt}{0.400pt}}
\put(181,123){\makebox(0,0)[r]{\tiny 0}}
\put(1419.0,123.0){\rule[-0.200pt]{4.818pt}{0.400pt}}
\put(201.0,228.0){\rule[-0.200pt]{4.818pt}{0.400pt}}
\put(181,228){\makebox(0,0)[r]{\tiny 0.05}}
\put(1419.0,228.0){\rule[-0.200pt]{4.818pt}{0.400pt}}
\put(201.0,334.0){\rule[-0.200pt]{4.818pt}{0.400pt}}
\put(181,334){\makebox(0,0)[r]{\tiny 0.1}}
\put(1419.0,334.0){\rule[-0.200pt]{4.818pt}{0.400pt}}
\put(201.0,439.0){\rule[-0.200pt]{4.818pt}{0.400pt}}
\put(181,439){\makebox(0,0)[r]{\tiny 0.15}}
\put(1419.0,439.0){\rule[-0.200pt]{4.818pt}{0.400pt}}
\put(201.0,544.0){\rule[-0.200pt]{4.818pt}{0.400pt}}
\put(181,544){\makebox(0,0)[r]{\tiny 0.2}}
\put(1419.0,544.0){\rule[-0.200pt]{4.818pt}{0.400pt}}
\put(201.0,649.0){\rule[-0.200pt]{4.818pt}{0.400pt}}
\put(181,649){\makebox(0,0)[r]{\tiny 0.25}}
\put(1419.0,649.0){\rule[-0.200pt]{4.818pt}{0.400pt}}
\put(201.0,755.0){\rule[-0.200pt]{4.818pt}{0.400pt}}
\put(181,755){\makebox(0,0)[r]{\tiny 0.3}}
\put(1419.0,755.0){\rule[-0.200pt]{4.818pt}{0.400pt}}
\put(201.0,860.0){\rule[-0.200pt]{4.818pt}{0.400pt}}
\put(181,860){\makebox(0,0)[r]{\tiny 0.35}}
\put(1419.0,860.0){\rule[-0.200pt]{4.818pt}{0.400pt}}
\put(201.0,123.0){\rule[-0.200pt]{0.400pt}{4.818pt}}
\put(201,82){\makebox(0,0){\tiny 0}}
\put(201.0,840.0){\rule[-0.200pt]{0.400pt}{4.818pt}}
\put(449.0,123.0){\rule[-0.200pt]{0.400pt}{4.818pt}}
\put(449,82){\makebox(0,0){\tiny 2}}
\put(449.0,840.0){\rule[-0.200pt]{0.400pt}{4.818pt}}
\put(696.0,123.0){\rule[-0.200pt]{0.400pt}{4.818pt}}
\put(696,82){\makebox(0,0){\tiny 4}}
\put(696.0,840.0){\rule[-0.200pt]{0.400pt}{4.818pt}}
\put(944.0,123.0){\rule[-0.200pt]{0.400pt}{4.818pt}}
\put(944,82){\makebox(0,0){\tiny 6}}
\put(944.0,840.0){\rule[-0.200pt]{0.400pt}{4.818pt}}
\put(1191.0,123.0){\rule[-0.200pt]{0.400pt}{4.818pt}}
\put(1191,82){\makebox(0,0){\tiny 8}}
\put(1191.0,840.0){\rule[-0.200pt]{0.400pt}{4.818pt}}
\put(1439.0,123.0){\rule[-0.200pt]{0.400pt}{4.818pt}}
\put(1439,82){\makebox(0,0){\tiny 10}}
\put(1439.0,840.0){\rule[-0.200pt]{0.400pt}{4.818pt}}
\put(40,491){\makebox(0,0){\tiny PL($\ell$)}}
\put(820,21){\makebox(0,0){\tiny $\ell$}}
\put(1224,600){\makebox(0,0)[r]{\tiny E. Coli PPI}}
\put(1244.0,600.0){\rule[-0.200pt]{11.090pt}{0.400pt}}
\put(201,178){\usebox{\plotpoint}}
\multiput(201.58,178.00)(0.499,0.936){245}{\rule{0.120pt}{0.848pt}}
\multiput(200.17,178.00)(124.000,230.239){2}{\rule{0.400pt}{0.424pt}}
\multiput(325.58,410.00)(0.499,1.074){245}{\rule{0.120pt}{0.958pt}}
\multiput(324.17,410.00)(124.000,264.011){2}{\rule{0.400pt}{0.479pt}}
\multiput(449.00,676.58)(0.549,0.499){221}{\rule{0.539pt}{0.120pt}}
\multiput(449.00,675.17)(121.881,112.000){2}{\rule{0.270pt}{0.400pt}}
\multiput(572.58,783.09)(0.499,-1.357){245}{\rule{0.120pt}{1.184pt}}
\multiput(571.17,785.54)(124.000,-333.543){2}{\rule{0.400pt}{0.592pt}}
\multiput(696.58,448.96)(0.499,-0.791){245}{\rule{0.120pt}{0.732pt}}
\multiput(695.17,450.48)(124.000,-194.480){2}{\rule{0.400pt}{0.366pt}}
\multiput(820.00,254.92)(0.796,-0.499){153}{\rule{0.736pt}{0.120pt}}
\multiput(820.00,255.17)(122.473,-78.000){2}{\rule{0.368pt}{0.400pt}}
\multiput(944.00,176.92)(1.599,-0.498){75}{\rule{1.372pt}{0.120pt}}
\multiput(944.00,177.17)(121.153,-39.000){2}{\rule{0.686pt}{0.400pt}}
\multiput(1068.00,137.92)(4.858,-0.493){23}{\rule{3.885pt}{0.119pt}}
\multiput(1068.00,138.17)(114.937,-13.000){2}{\rule{1.942pt}{0.400pt}}
\put(1191,124.17){\rule{24.900pt}{0.400pt}}
\multiput(1191.00,125.17)(72.319,-2.000){2}{\rule{12.450pt}{0.400pt}}
\put(1220,650){\makebox(0,0)[r]{\tiny SN}}
\multiput(1244,650)(20.756,0.000){5}{\usebox{\plotpoint}}
\put(201,176){\usebox{\plotpoint}}
\multiput(201,176)(5.274,20.074){24}{\usebox{\plotpoint}}
\multiput(325,648)(15.096,14.244){8}{\usebox{\plotpoint}}
\multiput(449,765)(11.238,-17.450){11}{\usebox{\plotpoint}}
\multiput(572,574)(10.858,-17.689){12}{\usebox{\plotpoint}}
\multiput(696,372)(14.154,-15.181){8}{\usebox{\plotpoint}}
\multiput(820,239)(17.760,-10.742){7}{\usebox{\plotpoint}}
\multiput(944,164)(20.280,-4.416){6}{\usebox{\plotpoint}}
\multiput(1068,137)(20.673,-1.849){6}{\usebox{\plotpoint}}
\multiput(1191,126)(20.753,-0.335){6}{\usebox{\plotpoint}}
\multiput(1315,124)(20.755,-0.167){6}{\usebox{\plotpoint}}
\put(1439,123){\usebox{\plotpoint}}
\put(201.0,123.0){\rule[-0.200pt]{138.234pt}{0.400pt}}
\put(1439.0,123.0){\rule[-0.200pt]{0.400pt}{83.543pt}}
\put(201.0,860.0){\rule[-0.200pt]{138.234pt}{0.400pt}}
\put(201.0,123.0){\rule[-0.200pt]{0.400pt}{83.543pt}}
\end{picture}
\caption{Path length distribution of the given {\it E. Coli} PPI network and a typical outcome of the SN model (SN).}
\label{fig:MRSAlength}
\end{center}
\end{figure}

The motif distribution of the generated networks having at most 20$\%$ discrepancy from the average results is equal to the one of the give network.
The clustering coefficient distribution of the given network has a trend $\gamma = -0.52$ while similar trend for the generated networks having at most 20$\%$ discrepancy from the average results is $\gamma = -0.47$.
The path length distribution of the given network and of one of the generated networks having at most 20$\%$ discrepancy from the average results is depicted in \fig{fig:MRSAlength}.

\section{Final remarks}
\label{sec:fr}
In this section we give some thoughts about the SN model and we suggest possible directions for research on this model.

It seems that the SN model shows that preferential attachment is not necessary to generate networks having a power law degree distribution.
Can it be the preferential attachment is somehow `hidden' in the SN model?
We think that preferential attachment is `hidden' in the combination of structured nodes and Hamming distance.
Anyhow, it is surprising that these two simple concepts (Hamming distance is rather simple when compared to the reasons behind the interactions in gene and protein networks) can behave as preferential attachment and, in some cases, (as for the average clustering coefficient being independent from the network size or the high heterogeneity index) be better in reproducing empirical networks.

The study on the SN model is still in its very early stages in order to allow us to say something new about biological networks.
We do not think that the SN model can recreate all empirical networks or all features of some empirical network, anyhow, it is interesting to note that this model can recreate a broad range of topological features present in empirical networks of different nature.
Of course, the big number of input parameters (and their domain) of the SN model allows to `tune' some of the features of the generated networks more than what possible with other network models.

As we said, we used a `trial-and-error' approach in order to find the instances of the SN model generating the networks considered by us. 
For some of these networks we got pretty close results, for others less good results.
We did not aim to have an exact match for the empirical networks considered by us, we aimed to show the broad range of topological features that can be matched by the SN model.

It is definitely interesting to study the classes of networks that can be generated by the SN model upon changes in its input parameters.
Moreover, extensions to the model will allow it to generate directed networks, to evolve networks or to use the generated networks for other studies.
For instance, one might need to have networks with a specific motif distribution in order to study their dynamics.\\

\noindent
{\bf Acknowledgements}
We gratefully acknowledge Ian Overton for providing the MRSA network.

\bibliographystyle{plain}

\end{document}